\documentclass[12pt]{article}
\usepackage{amssymb,amscd,array}
\catcode `\@=11
\@addtoreset{equation}{section}

\newtheorem{thm}{Theorem}[section]

\def\qed{\blacksquare}
\newcommand{\be}{\begin{equation}}
\newcommand{\ee}{\end{equation}}
\newcommand{\bea}{\begin{eqnarray}}
\newcommand{\eea}{\end{eqnarray}}

\newcommand{\N}{\mathbb{N}}

\begin{document}
\begin{titlepage}

\begin{center}
{\bf \Large{Anti-BRST in the Causal Approach\\}}
\end{center}
\vskip 1.0truecm
\centerline{D. R. Grigore, 
\footnote{e-mail: grigore@theory.nipne.ro}}
\vskip5mm
\centerline{Department of Theoretical Physics}
\centerline{ Institute for Physics and Nuclear Engineering ``Horia Hulubei"}
\centerline{Bucharest-M\u agurele, P. O. Box MG 6, ROM\^ANIA}

\vskip 2cm
\bigskip \nopagebreak
\begin{abstract}
\noindent
It is known that the elimination of the anomalies in all orders of perturbation theory is an open problem. The constrains
given by usual invariance properties and the Wess-Zumino identities are not enough to eliminate the anomalies in the 
general case of an Yang-Mills theory. So, any new symmetry of the model could restrict further the anomalies and be a
solution of the problem. 

We consider the anti-BRST transform of Ojima in the causal approach and investigate if such new restrictions are obtained.
Unfortunately, the result is negative: if we have BRST invariance up to the second order of the perturbation theory, we also
have anti-BRST invariance up to the same order. Probably, this result is true in all orders of the perturbation theory.

So, anti-BRST transform gives nothing new, and we have to find other ideas to restrict and eventually eliminate the anomalies
for a general Yang-Mills theory.

\end{abstract}
\end{titlepage}

\section{Introduction}

The general framework of perturbation theory consists in the construction of
some
distribution-valued operators called chronological products \cite{BS}. We prefer
the 
framework from \cite{DF}: for every set of Wick monomials 
$ 
W_{1}(x_{1}),\dots,W_{n}(x_{n}) 
$
acting in some Fock space
$
{\cal H}
$
one associates the distribution-valued operator
$ 
T(W_{1}(x_{1}),\dots,W_{n}(x_{n})) \equiv
T^{W_{1},\dots,W_{n}}(x_{1},\dots,x_{n})
$
such that a set of axioms, essentially proposed by Bogoliubov, are verified.
The modern construction of the chronological products can be done recursively
according to Epstein-Glaser prescription \cite{EG}, \cite{Gl}, \cite{Sto1}, \cite{Sc1}, \cite{Sc2} (which reduces the induction
procedure to a distribution splitting of some distributions with causal support)
or according to Stora prescription \cite{PS} (which reduces the renormalization
procedure to the process of extension of distributions). These products are not
uniquely defined but there are some natural limitation on the arbitrariness. If
the arbitrariness does not grow with $n$ (the order of the perturbation theory) we have a renormalizable theory. An
equivalent point of view uses retarded products \cite{St1}.

The description of higher spins in the perturbation theory can be problematic.
If we describe them by fields carrying only physical degrees of freedom, then 
the theories are usually not renormalizable. However, one can save renormalizability 
using ghost fields. Such theories are defined in a Fock space
$
{\cal H}
$
with indefinite metric, generated by physical and un-physical fields (called
{\it ghost fields}). One selects the physical states assuming the existence of
an operator $Q$ called {\it gauge charge} which verifies
$
Q^{2} = 0
$
and such that the {\it physical Hilbert space} is by definition
$
{\cal H}_{\rm phys} \equiv Ker(Q)/Im(Q).
$
The fact that two distinct mathematical states from 
$
{\cal H}
$
can be associated to the same physical context is called {\it gauge freedom} and the 
corresponding theories are called {\it gauge theories}.
The graded commutator
$
d_{Q}
$
of the gauge charge with any operator $A$ of fixed ghost number
\be
d_{Q}A = [Q,A]
\ee
(where
$
[\cdot,\cdot]
$
denotes the graded commutator) verifies
\be
d_{Q}^{2} = 0
\ee
so
$
d_{Q}
$
is a co-chain operator in the space of Wick polynomials. 
 
A gauge theory assumes also that there exists a Wick polynomial of null ghost
number
$
T(x)
$
called {\it the interaction Lagrangian} such that
\be
d_{Q}T = i \partial_{\mu}T^{\mu}
\label{gau1}
\ee
for some other Wick polynomials
$
T^{\mu}.
$
This relation means that the expression $T$ leaves invariant the physical
states, at least in the adiabatic limit. Indeed, we have:
\be
T(f)~{\cal H}_{\rm phys}~\subset~~{\cal H}_{\rm phys}  
\label{gau2}
\ee
up to terms which can be made as small as desired (making the test function $f$
flatter and flatter). In all known models one finds out that there exist a chain
of Wick polynomials
$
T^{\mu},~T^{[\mu\nu]},~T^{[\mu\nu\rho]},\dots
$
such that:
\be
d_{Q}T = i \partial_{\mu}T^{\mu}, \quad
d_{Q}T^{\mu} = i \partial_{\nu}T^{[\mu\nu]}, \quad
d_{Q}T^{[\mu\nu]} = i \partial_{\rho}T^{[\mu\nu\rho]},\dots
\label{descent}
\ee
where the brackets emphasize completely antisymmetric in all indexes; it follows that 
the chain of relation stops after a finite number of steps. We can also use
a compact notation
$
T^{I}
$
where $I$ is a collection of indexes
$
I = [\nu_{1},\dots,\nu_{p}]~(p = 0,1,\dots,)
$
and one can write compactly the relations (\ref{descent}) as follows:
\be
d_{Q}T^{I} = i~\partial_{\mu}T^{I\mu}.
\label{descent1}
\ee
All these polynomials have the same canonical dimension
\be
\omega(T^{I}) = \omega_{0},~\forall I
\ee
and the ghost number:
\be
gh(T^{I}) = |I|.
\ee
If the interaction Lagrangian $T$ is Lorentz invariant, then one can prove that
the expressions
$
T^{I},~|I| > 0
$
can be taken Lorentz covariant.

Now we can construct the chronological products
\be
T^{I_{1},\dots,I_{n}}(x_{1},\dots,x_{n}) \equiv
T(T^{I_{1}}(x_{1}),\dots,T^{I_{n}}(x_{n}))
\ee
according to the recursive procedure. We say that the theory is gauge invariant
in all orders of the perturbation theory if the following set of identities
generalizing (\ref{descent1}):
\be
d_{Q}T^{I_{1},\dots,I_{n}} = 
i \sum_{l=1}^{n} (-1)^{s_{l}} {\partial\over \partial x^{\mu}_{l}}
T^{I_{1},\dots,I_{l}\mu,\dots,I_{n}}
\label{gauge}
\ee
are true for all 
$n \in \N$
and all
$
I_{1}, \dots, I_{n}.
$
Here we have defined
\be
s_{l} \equiv \sum_{j=1}^{l-1} |I|_{j}.
\ee

Such identities can be usually broken by {\it anomalies} i.e. expressions of the
type
$
A^{I_{1},\dots,I_{n}}
$
which are quasi-local and might appear in the right-hand side of the relation
(\ref{gauge}). These anomalies are constrained by some identities (the Wess-Zumino relations).
It still an unsolved problem to prove that the anomalies can be eliminated by convenient redefinitions
of the chronological products. This problem is not resolved satisfactory in other approaches also, at least 
in our opinion.

One idea to eliminate the anomalies is to try to find new restrictions verified by them beside
the Wess-Zumino relations. One possibility could be the anti-BRST transform introduced in \cite{O}.
We will prove in this paper a negative result, namely that up to the second order of perturbation
theory, the anti-BRST transform does not produce new constrains on the model; that is if we impose
BRST symmetry up to the second order, we automatically have anti-BRST up to the second order.

In the next Section we will briefly present the Yang-Mills model in our preferred compact
notations. In Section \ref{anti} we prove our main result. 

\newpage
\section{Yang-Mills Models in the Causal Formalism\label{ym}}
We give some results from \cite{cohomology}.
\subsection{Massless Particles of Spin $1$ (Photons)}

We consider a vector space 
$
{\cal H}
$
of Fock type generated (in the sense of Borchers theorem) by the vector field 
$
v_{\mu}
$ 
(with Bose statistics) and the scalar fields 
$
u, \tilde{u}
$
(with Fermi statistics). The Fermi fields are usually called {\it ghost fields}.
We suppose that all these (quantum) fields are of null mass. Let $\Omega$ be the
vacuum state in
$
{\cal H}.
$
In this vector space we can define a sesquilinear form 
$<\cdot,\cdot>$
in the following way: the (non-zero) $2$-point functions are by definition:
\bea
<\Omega, v_{\mu}(x_{1}) v_{\mu}(x_{2})\Omega> =i~\eta_{\mu\nu}~D_{0}^{(+)}(x_{1}
- x_{2}),
\nonumber \\
<\Omega, u(x_{1}) \tilde{u}(x_{2})\Omega> =- i~D_{0}^{(+)}(x_{1} - x_{2})
\qquad
<\Omega, \tilde{u}(x_{1}) u(x_{2})\Omega> = i~D_{0}^{(+)}(x_{1} - x_{2})
\eea
and the $n$-point functions are generated according to Wick theorem. Here
$
\eta_{\mu\nu}
$
is the Minkowski metrics (with diagonal $1, -1, -1, -1$) and 
$
D_{0}^{(+)}
$
is the positive frequency part of the Pauli-Jordan distribution
$
D_{0}
$
of null mass. To extend the sesquilinear form to
$
{\cal H}
$
we define the conjugation by
\be
v_{\mu}^{\dagger} = v_{\mu}, \qquad 
u^{\dagger} = u, \qquad
\tilde{u}^{\dagger} = - \tilde{u}.
\ee

Now we can define in 
$
{\cal H}
$
the operator $Q$ according to the following formulas:
\bea
~[Q, v_{\mu}] = i~\partial_{\mu}u,\qquad
[Q, u] = 0,\qquad
[Q, \tilde{u}] = - i~\partial_{\mu}v^{\mu}
\nonumber \\
Q\Omega = 0
\label{Q-0}
\eea
where by 
$
[\cdot,\cdot]
$
we mean the graded commutator. One can prove that $Q$ is well defined: basically it leaves
invariant the causal commutation relations.  The usefulness of this construction follows 
from:
\begin{thm}
The operator $Q$ verifies
$
Q^{2} = 0.
$ 
The factor space
$
Ker(Q)/Ran(Q)
$
is isomorphic to the Fock space of particles of zero mass and helicity $1$
(photons). 
\end{thm}


\subsection{Massive Particles of Spin $1$ (Heavy Bosons)}

We repeat the whole argument for the case of massive photons i.e. particles of
spin $1$ and positive mass. 

We consider a vector space 
$
{\cal H}
$
of Fock type generated by the vector field 
$
v_{\mu},
$ 
the scalar field 
$
\Phi
$
(with Bose statistics) and the scalar fields 
$
u, \tilde{u}
$
(with Fermi statistics). We suppose that all these (quantum) fields are of mass
$
m > 0.
$
In this vector space we can define a sesquilinear form 
$<\cdot,\cdot>$
in the following way: the (non-zero) $2$-point functions are by definition:
\bea
<\Omega, v_{\mu}(x_{1}) v_{\mu}(x_{2})\Omega> =i~\eta_{\mu\nu}~D_{m}^{(+)}(x_{1}
- x_{2}),
\quad
<\Omega, \Phi(x_{1}) \Phi(x_{2})\Omega> =- i~D_{m}^{(+)}(x_{1} - x_{2})
\nonumber \\
<\Omega, u(x_{1}) \tilde{u}(x_{2})\Omega> =- i~D_{m}^{(+)}(x_{1} - x_{2}),
\qquad
<\Omega, \tilde{u}(x_{1}) u(x_{2})\Omega> = i~D_{m}^{(+)}(x_{1} - x_{2})
\eea
and the $n$-point functions are generated according to Wick theorem. Here
$
D_{m}^{(+)}
$
is the positive frequency part of the Pauli-Jordan distribution
$
D_{m}
$
of mass $m$. To extend the sesquilinear form to
$
{\cal H}
$
we define the conjugation by
\be
v_{\mu}^{\dagger} = v_{\mu}, \qquad 
u^{\dagger} = u, \qquad
\tilde{u}^{\dagger} = - \tilde{u},
\qquad \Phi^{\dagger} = \Phi.
\ee

Now we can define in 
$
{\cal H}
$
the operator $Q$ according to the following formulas:
\bea
~[Q, v_{\mu}] = i~\partial_{\mu}u,\qquad
[Q, u] = 0,\qquad
[Q, \tilde{u}] = - i~(\partial_{\mu}v^{\mu} + m~\Phi)
\qquad
[Q,\Phi] = i~m~u,
\nonumber \\
Q\Omega = 0.
\label{Q-m}
\eea
One can prove that $Q$ is well defined. We have a result similar to the first
theorem of this Section:
\begin{thm}
The operator $Q$ verifies
$
Q^{2} = 0.
$ 
The factor space
$
Ker(Q)/Ran(Q)
$
is isomorphic to the Fock space of particles of mass $m$ and spin $1$ (massive
photons). 
\end{thm}
\subsection{The Generic Yang-Mills Case}

The situations described above (of massless and massive photons) are susceptible
of the following generalizations. We can consider a system of particles of null mass and helicity $1$ using the 
idea of the first Subsection for triplets
$
(v^{\mu}_{a}, u_{a}, \tilde{u}_{a}), a \in I_{1}
$
of massless fields; here
$
I_{1}
$
is a set of indexes. All the relations have to be modified by appending an index $a$ to all these
fields. 

In the massive case we have to consider quadruples
$
(v^{\mu}_{a}, u_{a}, \tilde{u}_{a}, \Phi_{a}),  a \in I_{2}
$
of fields of mass 
$
m_{a}
$
and use the second Subsection; here
$
I_{2}
$
is another set of indexes.

We can consider now the most general case involving fields of spin not greater
that $1$.
We take 
$
I = I_{1} \cup I_{2} \cup I_{3}
$
a set of indexes and for any index we take a quadruple
$
(v^{\mu}_{a}, u_{a}, \tilde{u}_{a},\Phi_{a}), a \in I
$
of fields with the following conventions:
(a) For
$
a \in I_{1}
$
we impose 
$
\Phi_{a} = 0
$
and we take the masses to be null
$
m_{a} = 0;
$
(b) For
$
a \in I_{2}
$
we take the all the masses strictly positive:
$
m_{a} > 0;
$
(c) For 
$
a \in I_{3}
$
we take 
$
v_{a}^{\mu}, u_{a}, \tilde{u}_{a}
$
to be null and the fields
$
\Phi_{a} \equiv \phi^{H}_{a} 
$
of mass 
$
m^{H}_{a} \geq 0.
$
The fields
$
\phi^{H}_{a} 
$
are called {\it Higgs fields}.

If we define
$
m_{a} = 0, \forall a \in I_{3}
$
then we can define in 
$
{\cal H}
$
the operator $Q$ according to the following formulas for all indexes
$
a \in I:
$
\bea
~[Q, v^{\mu}_{a}] = i~\partial^{\mu}u_{a},\qquad
[Q, u_{a}] = 0,
\nonumber \\
~[Q, \tilde{u}_{a}] = - i~(\partial_{\mu}v^{\mu}_{a} + m_{a}~\Phi_{a})
\qquad
[Q,\Phi_{a}] = i~m_{a}~u_{a},
\nonumber \\
Q\Omega = 0.
\label{Q-general}
\eea

If we consider matter fields also i.e some set of Dirac fields with Fermi
statistics:
$
\psi_{A}, A \in I_{4}
$ 
then we impose
\be
d_{Q}\psi_{A} = 0.
\ee

\newpage
\subsection{The Yang-Mills Interaction\label{interaction}}

In the framework and notations from the end of the preceding Subsection we have the 
following result which describes the most general form of the Yang-Mills interaction
\cite{YM}, \cite{standard}, \cite{fermi}. Summation over the dummy indexes is used 
everywhere.
\begin{thm}
Let $T$ be a relative cocycle for 
$
d_{Q}
$
i.e. verifies (\ref{gau1}) and also: 
1) is tri-linear in the fields; 2) is of canonical dimension
$
\omega(T) \leq 4;
$
3) has ghost number
$
gh(T) = 0.
$
Then:

(i) $T$ is (relatively) cohomologous to a non-trivial co-cycle of the form:
\bea
T = f_{abc} \left( {1\over 2}~v_{a\mu}~v_{b\nu}~F_{c}^{\nu\mu}
+ u_{a}~v_{b}^{\mu}~\partial_{\mu}\tilde{u}_{c}\right)
\nonumber \\
+ f^{\prime}_{abc} (\Phi_{a}~\partial_{\mu}\Phi_{b}~v_{c}^{\mu} - m_{b}~\Phi_{a}~v_{b}^{\mu}~v_{c\mu}
+ m_{b}~\Phi_{a}~\tilde{u}_{b}~u_{c})
\nonumber \\
+ {1\over 3!}~f^{\prime\prime}_{abc}~\Phi_{a}~\Phi_{b}~\Phi_{c}
+ j^{\mu}_{a}~v_{a\mu} + j_{a}~\Phi_{a};
\label{T}
\eea
where we can take the constants
$
f_{abc} = 0
$
if one of the indexes is in
$
I_{3};
$
also
$
f^{\prime}_{abc} = 0
$
if 
$
c \in I_{3}
$
or one of the indexes $a$ and $b$ are from
$
I_{1};
$
and
$
j^{\mu}_{a} = 0
$
if
$
a \in I_{3};
$
$
j_{a} = 0
$
if
$
a \in I_{1}.
$
Moreover we have:

(a) The constants
$
f_{abc}
$
are completely antisymmetric
\be
f_{abc} = f_{[abc]}.
\label{anti-f}
\ee

(b) The expressions
$
f^{\prime}_{abc}
$
are antisymmetric  in the indexes $a$ and $b$:
\be
f^{\prime}_{abc} = - f^{\prime}_{bac}
\label{anti-f'}
\ee
and are connected to 
$f_{abc}$
by:
\be
f_{abc}~m_{c} = f^{\prime}_{cab} m_{a} - f^{\prime}_{cba} m_{b}.
\label{f-f'}
\ee

(c) The (completely symmetric) expressions 
$f^{\prime\prime}_{abc} = f^{\prime\prime}_{\{abc\}}$
verify
\be
f^{\prime\prime}_{abc} = \left\{\begin{array}{rcl} 
{1 \over m_{c}}~f'_{abc}~(m_{a}^{2} - m_{b}^{2}) & \mbox{for} & a, b \in I_{3},
c \in I_{2} \\
- {1 \over m_{c}}~f'_{abc}~m_{b}^{2} & \mbox{for} & a, c \in I_{2}, b \in
I_{3}.\end{array}\right.
\label{f"}
\ee

(d) the expressions
$
j^{\mu}_{a}
$
and
$
j_{a}
$
are bilinear in the Fermi matter fields: in tensor notations;
\bea
j_{a}^{\mu} = \sum_{\epsilon}~
\overline{\psi} t^{\epsilon}_{a} \otimes \gamma^{\mu}\gamma_{\epsilon} \psi
\qquad
j_{a} = \sum_{\epsilon}~
\overline{\psi} s^{\epsilon}_{a} \otimes \gamma_{\epsilon} \psi
\label{current}
\eea
where  for every
$
\epsilon = \pm
$
we have defined the chiral projectors of the algebra of Dirac matrices
$
\gamma_{\epsilon} \equiv {1\over 2}~(I + \epsilon~\gamma_{5})
$
and
$
t^{\epsilon}_{a},~s^{\epsilon}_{a}
$
are 
$
|I_{4}| \times |I_{4}|
$
matrices. If $M$ is the mass matrix
$
M_{AB} = \delta_{AB}~M_{A}
$
then we must have
\be
\partial_{\mu}j^{\mu}_{a} = m_{a}~j_{a} 
\qquad \Leftrightarrow \qquad
m_{a}~s_{a}^{\epsilon} = i(M~t^{\epsilon}_{a} - t^{-\epsilon}_{a}~M).
\label{conserved-current}
\ee

(ii) The relation 
$
d_{Q}T = i~\partial_{\mu}T^{\mu}
$
is verified by:
\be
T^{\mu} = f_{abc} \left( u_{a}~v_{b\nu}~F^{\nu\mu}_{c} -
{1\over 2} u_{a}~u_{b}~\partial^{\mu}\tilde{u}_{c} \right)
+ f^{\prime}_{abc}~(\Phi_{a}~\partial^{\mu}\Phi_{b}~u_{c} - m_{b}~\Phi_{a}~v^{\mu}_{b}~u_{c})
+ j^{\mu}_{a}~u_{a}
\label{Tmu}
\ee

(iii) The relation 
$
d_{Q}T^{\mu} = i~\partial_{\nu}T^{\mu\nu}
$
is verified by:
\be
T^{\mu\nu} \equiv {1\over 2} f_{abc}~u_{a}~u_{b}~F_{c}^{\mu\nu}.
\ee
\label{T1}
\end{thm}

Now if we impose gauge invariance in the second order of perturbation theory i.e. (\ref{gauge}) for 
$n = 2$
we get new constrains on the constants from the interaction Lagrangian (for an extensive treatment see \cite{YM} - \cite{cohomology}):

\be
\sum_{c}~(f_{abc}~f_{dec} + f_{bdc}~f_{aec} + f_{dac}~f_{bec}) = 0
\label{Jacobi}
\ee
(Jacobi identity)
\be
\sum_{c}~[ f^{\prime}_{dca}~f^{\prime}_{ceb}
- (a \leftrightarrow b) ] = 
- \sum_{c}~f_{abc}~f^{\prime}_{dec},
\qquad
a,b \in I_{1} \cup I_{2},~d,e \in I_{2} \cup I_{3}
\label{reps1}
\ee
(the representation property for the Bose sector) 
\be
[ t_{a}^{\epsilon}, t_{b}^{\epsilon} ] = i~f_{abc}~t_{c}^{\epsilon}
\label{reps2}
\ee
(the representation property for the Fermi sector)
\be
t_{a}^{- \epsilon}~s_{b}^{\epsilon} - s_{b}^{\epsilon}~t_{a}^{\epsilon}
= i~f^{\prime}_{bca}~s_{c}^{\epsilon} 
\label{tensor}
\ee
(a tensor representation property for the expressions $s_{a}$). 
\newpage
\subsection{Causal Perturbation Theory\label{causal}}
 
We give the idea of the construction of the chronological products only in the second order of the perturbation theory, relevant 
for our result. The basic construction of Epstein and Glaser is the construction of the causal
commutator. More details can be found in \cite{cohomology}. We mention the basic procedure used in the causal approach.
If we want to construct the chronological product
$
T(A(x_{1}),B(x_{2}))
$
for arbitrary Wick monomials $A$ and $B$, the idea of Epstein and Glaser is to consider the (graded) commutaror
\be
D(x_{1}, x_{2}) \equiv [A(x_{1}),B(x_{2})].
\ee
It is easy to see that the tree contribution to this commutator is of the form:
\be
D_{(0)}(x_{1}, x_{2}) = \sum p_{j}(\partial)D(x_{1} - x_{2})~W_{j}(x_{1}, x_{2}) 
\label{com-02}
\ee
where 
$
p_{j}
$ 
are polynomials in the partial derivatives and 
$
W_{j}
$
are Wick polynomials. Then one can obtain the associated chronological products if one makes in the preceding formula 
the substitution
$
D \rightarrow D^{F}
$
i.e. if we replace the Pauli-Villars causal distribution with the Feynman propagator:
\be
T_{(0)}(x_{1}, x_{2}) = \sum p_{j}(\partial)D^{F}(x_{1} - x_{2})~W_{j}(x_{1}, x_{2}). 
\label{chr-02}
\ee
In this way we fulfill Bogoliubov axioms in the second order for the tree contributions. However, in this way we can produce
anomalies in the relation (\ref{gauge}). The reason is that for the gauge invariance of the causal commutators - e.g. relation (\ref{1}) 
from the next Section - we need the Klein-Gordon equation for the Pauli-Villars distribution
$
(\square + m^{2})~D_{m} = 0
$
but for the gauge invariance of the chronological products we must use 
$
(\square + m^{2})~D^{F}_{m} = \delta.
$
In some cases it is possible to eliminate these anomalies if we use finite 
renormalizations
\be
T_{(0)}(x_{1}, x_{2}) \rightarrow T^{\rm ren}_{(0)}(x_{1}, x_{2}) \equiv T_{(0)}(x_{1}, x_{2}) + N(x_{1}, x_{2})
\label{Renormalisation}
\ee
where 
$N$ are finite renormalizations; they must be quasi-local expressions i. e they are of the form 
$
q_{j}(\partial)\delta(x_{1} - x_{2})~N_{j}(x_{1})
$
with 
$
N_{j}
$ 
Wick polynomials. This program can be extended, in principle to loop 
contributions and to higher orders of the perturbation theory, but a complete analysis is not available now.
\newpage
\section{Anti-BRST Transform\label{anti}}

In Section \ref{ym} we have introduced the BRST transform; we prefer to rewrite it using the (graded) commutator
$
d_{Q}
$:
\bea
d_{Q}v_{a}^{\mu} = i~\partial^{\mu}u_{a},\qquad
d_{Q}u_{a} = 0,\qquad
d_{Q} \tilde{u}_{a} = - i~(\partial_{\mu}v^{\mu}_{a} + m~\Phi_{a}),
\qquad
d_{Q}\Phi_{a} = i~m_{a}~u_{a},
\nonumber\\
d_{Q}\psi_{A} = 0.
\label{brs}
\eea

These expressions are the linear part of the non-linear BRST transform from the classical field theory. We proceed in the same 
spirit with the anti-BRST transform from \cite{O} and obtain the operator
$
d^{\rm anti}_{Q}
$
(associated to the anti-BRST transform
$
Q_{\rm anti})
$
and given by:
\bea
d^{\rm anti}_{Q}v_{a}^{\mu} = i~\partial^{\mu}\tilde{u}_{a},\qquad
d^{\rm anti}_{Q}u_{a} = i~(\partial_{\mu}v^{\mu}_{a} + m~\Phi_{a}),\qquad
d^{\rm anti}_{Q} \tilde{u}_{a} = 0,
\qquad
d^{\rm anti}_{Q}\Phi_{a} = i~m_{a}~\tilde{u}_{a},
\nonumber\\
d^{\rm anti}_{Q}\psi_{A} = 0,
\label{anti-brs}
\eea
and
\be
Q_{\rm anti} \Omega = 0.
\ee

As in the case of the BRST transform (\ref{brs}), we immediately check that
\be
(Q_{\rm anti})^{2} = 0 \qquad \Longleftrightarrow  \qquad (d^{\rm anti}_{Q})^{2} = 0.
\ee

We also have
\be
\{ Q, Q_{\rm anti} \} = 0.
\ee

Now we want to investigate if a theorem similar to \ref{T1} is valid for the anti-BRST transform. Indeed we have:

\begin{thm}
Suppose that the conditions from theorem \ref{T1} are fulfilled. Then:

(i) The relation 
$
d^{\rm anti}_{Q}T = i~\partial_{\mu}T_{\rm anti}^{\mu}
$
is verified for 
\bea
T_{\rm anti}^{\mu} = f_{abc} \left( \tilde{u}_{a}~v_{b\nu}~F^{\nu\mu}_{c} + \tilde{u}_{a}~\partial_{\nu}v_{b}^{\nu}~v^{\mu}_{c}
- u_{a}~\tilde{u}_{b}~\partial^{\mu}\tilde{u}_{c}
+ {1\over 2} \partial^{\mu}u_{a}~\tilde{u}_{b}~\tilde{u}_{c} - m_{c}~\tilde{u}_{a}~v_{b}^{\mu}~\Phi_{c}\right)
\nonumber\\
+ f^{\prime}_{abc}~\left( \Phi_{a}~\partial^{\mu}\Phi_{b}~\tilde{u}_{c} - m_{b}~\Phi_{a}~v_{b}^{\mu}~\tilde{u}_{c}\right)
+ j^{\mu}_{a}~\tilde{u}_{a}
\label{Tmu-anti}
\eea

(ii) The relation 
$
d^{\rm anti}_{Q}T_{\rm anti}^{\mu} = i~\partial_{\nu}T_{\rm anti}^{\mu\nu}
$
is verified by:
\be
T_{\rm anti}^{\mu\nu} \equiv {1\over 2} f_{abc}~\tilde{u}_{a}~\tilde{u}_{b}~F_{c}^{\mu\nu}.
\ee
\label{T2}
\end{thm}

{\bf Proof:} It is elementary, by direct computations. We must cleverly use all the linear relations between the 
constants of $T$ derived in theorem \ref{T1}.
$\qed$

We remark that the expression
$
T_{\rm anti}^{\mu\nu}
$
can be obtained from
$
T^{\mu\nu}
$
if we make the transformation
\be
u_{a} \longleftrightarrow - \tilde{u}_{a}
\label{uu}
\ee
which also preserves the canonical anti-commutation relations and maps $Q$ in 
$
Q_{\rm anti};
$
however this is not true for 
$
T^{\mu} \longleftrightarrow T_{\rm anti}^{\mu}.
$

The preceding result shows that, as in \cite{O}, we get nothing new from anti-BRST in the first order of the perturbation theory.
We investigate now if the same is true for the second order of the perturbation theory. We are interested to impose gauge invariance 
for the anti-BRST transform, i.e. a relation of the same type as (\ref{gauge}) for the chronological products
\be
T_{\rm anti}^{I_{1},\dots,I_{n}}(x_{1},\dots,x_{n}) \equiv
T(T_{\rm anti}^{I_{1}}(x_{1}),\dots,T_{\rm anti}^{I_{n}}(x_{n})).
\ee
Such relations should be of the form:
\be
d^{\rm anti}_{Q} T(T_{\rm anti}^{I_{1}}(x_{1}),\dots,T_{\rm anti}^{I_{n}}(x_{n}))= 
i \sum_{l=1}^{n} (-1)^{s_{l}} {\partial\over \partial x^{\mu}_{l}}
T(T_{\rm anti}^{I_{1}}(x_{1}),\dots,T_{\rm anti}^{I_{l}\mu}(x_{l}),\dots,T_{\rm anti}^{I_{n}}(x_{n})).
\label{gauge-anti}
\ee
The hope would be to obtain new constrains on the possible anomalies. We have succeeded to study the tree contribution in the 
second order of perturbation theory, i.e. the preceding relation for
$
n = 2
$
and tree contributions. We have obtained the following result.

\begin{thm}
We can impose the relation
\bea
d^{\rm anti}_{Q}T_{(0)}(T_{\rm anti}^{I}(x_{1}), T_{\rm anti}^{J}(x_{2})) 
\nonumber\\
- i~{\partial \over \partial x^{\mu}_{1}}~T_{(0)}(T_{\rm anti}^{I\mu}(x_{1}), T_{\rm anti}^{J}(x_{2})) 
- i~(- 1)^{|I|}~{\partial \over \partial x^{\mu}_{2}}~T_{(0)}(T_{\rm anti}^{I}(x_{1}), T_{\rm anti}^{J\mu}(x_{2})) = 0
\label{anti-brst}
\eea
for the tree components of the chronological products if we perform the following finite renormalizations:
\be
T_{(0)}(T_{\rm anti}^{I}(x_{1}), T_{\rm anti}^{J}(x_{2})) \rightarrow 
T_{(0)}(T_{\rm anti}^{I}(x_{1}), T_{\rm anti}^{J}(x_{2})) - \delta(x_{1} - x_{2})~N^{IJ}(x_{1})
\ee
with the explicit expressions:
\be
N^{[\mu\nu]\emptyset} = {i \over 2}~f_{abe}~f_{cde}~\tilde{u}_{a}~\tilde{u}_{b}~v^{\mu}_{c}~v^{\nu}_{d}
\label{r1}
\ee
\be
N^{[\mu][\nu]} = - N^{[\mu\nu]\emptyset}
\label{r2}
\ee
\be
N^{[\mu]\emptyset} = g^{(1)}_{abcd}~\tilde{u}_{a}~v_{b}^{\mu}~v^{\nu}_{c}~v_{d\nu}
+ g^{(2)}_{abcd}~u_{a}~\tilde{u}_{b}~\tilde{u}_{c}~v_{d}^{\mu}
+ g^{(3)}_{abcd}~\tilde{u}_{a}~v_{b}^{\mu}~\Phi_{c}~\Phi_{d}
\label{r3}
\ee
with
\bea
g^{(1)}_{abcd} = - {i \over 2}~(f_{ace}~f_{bde} + f_{ade}~f_{bce})
\nonumber\\
g^{(2)}_{abcd} = - {i \over 2}~f_{ade}~f_{bce}
\nonumber\\
g^{(3)}_{abcd} = - {i \over 2}~(f^{\prime}_{cea}~f^{\prime}_{edb} + f^{\prime}_{dea}~f^{\prime}_{ecb})
\eea
and
\be
N^{\emptyset\emptyset} = h^{(1)}_{abcd}~v_{a\mu}~v_{b}^{\mu}~v^{\nu}_{c}~v_{d\nu}
+ h^{(2)}_{abcd}~v_{a}^{\mu}~v_{b\mu}~\Phi_{c}~\Phi_{d}
+ {1 \over 4!}~h^{(3)}_{abcd}~\Phi_{a}~\Phi_{b}~\Phi_{c}~\Phi_{d}
\label{r4}
\ee
with
\bea
h^{(1)}_{abcd} = - {i \over 4}~(f_{ade}~f_{bce} + f_{ace}~f_{bde})
\nonumber\\
h^{(2)}_{abcd} = - {i \over 2}~(f^{\prime}_{dea}~f^{\prime}_{ecb} + f^{\prime}_{cea}~f^{\prime}_{edb})
\nonumber\\
h^{(3)}_{abcd}~m_{a} = - 2~i~(f^{\prime}_{eba}~f^{\prime\prime}_{ecd} + f^{\prime}_{eca}~f^{\prime\prime}_{ebd}
+ f^{\prime}_{eda}~f^{\prime\prime}_{ebc}).
\eea
Only the linear and bilinear constrains from the end of the Subsection \ref{interaction} are needed for this result. 
The finite renormalization (\ref{r4}) is identical to the finite renormalization needed for the usual gauge invariance
of the tree contributions in the second order of the perturbation theory. 
\end{thm}
{\bf Proof:} (i) We start with the case 
$
I = [\mu\nu], J = \emptyset
$
of 
identity (\ref{gauge-anti}). We start from the identity
\be
d^{\rm anti}_{Q}[ T_{\rm anti}^{\mu\nu}(x_{1}), T(x_{2})) ] 
- i~{\partial \over \partial x^{\rho}_{2}}~[ T_{\rm anti}^{\mu\nu}(x_{1}), T_{\rm anti}^{\rho}(x_{2})] = 0
\label{1}
\ee
and do the substitution (\ref{com-02}) $\rightarrow$ (\ref{chr-02}). We need to collect all the terms from the 
commutator
$
[T_{\rm anti}^{\mu\nu}(x_{1}), T_{\rm anti}^{\rho}(x_{2})]
$
containing the factor
$
\partial^{\rho}D(x_{1} - x_{2}).
$
The identity (\ref{1}) holds if we use the Klein-Gordon equation for the Pauli-Villars distribution
$
(\square + m^{2})~D_{m} = 0.
$
However, when we make the substitution (\ref{com-02}) $\rightarrow$ (\ref{chr-02}) we must use 
$
(\square + m^{2})~D^{F}_{m} = \delta
$
and an anomaly appears. For instance, if we consider the first term of 
$
T_{\rm anti}^{\rho}(x_{2})
$
- see the expression (\ref{Tmu-anti}) - we have
\bea
[ T_{\rm anti}^{\mu\nu}(x_{1}), T_{\rm anti}^{\rho}(x_{2})_{1}] = 
{1 \over 2} f_{abc}~f_{pqr}~[ (\tilde{u}_{a}~\tilde{u}_{b}~F_{c}^{\mu\nu})(x_{1}), (\tilde{u}_{p}~v_{q\sigma}~F^{\sigma\rho}_{c})(x_{2})]
\nonumber\\
= {i \over 2}~f_{abe}~f_{cde}~
[ \partial^{\mu}\partial^{\rho}D_{m_{c}}(x_{1} - x_{2})~(\tilde{u}_{a}~\tilde{u}_{b})(x_{1}), (\tilde{u}_{c}~v_{d}^{\nu})(x_{2})
 - ( \mu \leftrightarrow \nu) ] + \cdots
 \label{11}
\eea
and after the substitution (\ref{com-02}) $\rightarrow$ (\ref{chr-02}) we obtain the anomaly
\be
A_{1}^{\mu\nu} = - {1 \over 2}~f_{abe}~f_{cde}~
[ \partial^{\mu}\delta(x_{1} - x_{2})~(\tilde{u}_{a}~\tilde{u}_{b})(x_{1})~(\tilde{u}_{c}~v_{d}^{\nu})(x_{2})
 - ( \mu \leftrightarrow \nu) ]
\ee
We have another term from the fourth term of 
$
T_{\rm anti}^{\rho}(x_{2})
$
and it is convenient to exhibit the end result in the form:
\be
A^{\mu\nu}(x_{1}, x_{2}) = \delta(x_{1} - x_{2})~a^{\mu\nu}(x_{1}) + \partial_{\rho}\delta(x_{2} - x_{1})~a^{\mu\nu;\rho}(x_{1}).
\ee
After some work, using Jacobi identity, we obtain that
\be
a^{\mu\nu;\rho} = 0
\qquad
a^{\mu\nu} = d_{Q}^{\rm anti}N^{[\mu\nu]\emptyset}
\ee
where 
$
N^{[\mu\nu]\emptyset}
$
is the expression from the statement. This anomaly can be eliminated by performing the finite renormalization (\ref{r1}).

(ii) Now we consider the identity
\be
d^{\rm anti}_{Q}[ T_{\rm anti}^{\mu}(x_{1}), T_{\rm anti}^{\nu}(x_{2})) ] 
- i~{\partial \over \partial x^{\rho}_{1}}~[ T_{\rm anti}^{\mu\rho}(x_{1}), T_{\rm anti}^{\nu}(x_{2})]
+ i~{\partial \over \partial x^{\rho}_{2}}~[ T_{\rm anti}^{\mu}(x_{1}), T_{\rm anti}^{\nu\rho}(x_{2}) ] = 0.
\label{2}
\ee
Now we need to collect all the terms from the two commutators
$
[T_{\rm anti}^{\mu\rho}(x_{1}), T_{\rm anti}^{\nu}(x_{2})]
$
and
$
[T_{\rm anti}^{\mu}(x_{1}), T_{\rm anti}^{\nu\rho}(x_{2})]
$
containing the factor
$
\partial^{\rho}D(x_{1} - x_{2});
$
afterwards we make the substitution (\ref{com-02}) $\rightarrow$ (\ref{chr-02}) and obtain the corresponding anomaly.
The computation leads, after using Jacobi identity, to the anomaly
\be
B^{\mu\nu}(x_{1}, x_{2}) = \delta(x_{1} - x_{2})~b^{\mu\nu}(x_{1})
\ee
where
\be
b^{\mu\nu} = d_{Q}^{\rm anti}N^{[\mu][\nu]}
\ee
with 
$
N^{[\mu][\nu]}
$
the expression from the statement. This anomaly can be eliminated by performing the finite renormalization (\ref{r2}).

(iii) The next step is the identity
\be
d^{\rm anti}_{Q}[ T_{\rm anti}^{\mu}(x_{1}), T(x_{2})) ] 
- i~{\partial \over \partial x^{\nu}_{1}}~[ T_{\rm anti}^{\mu\nu}(x_{1}), T(x_{2})]
+ i~{\partial \over \partial x^{\nu}_{2}}~[ T_{\rm anti}^{\mu}(x_{1}), T_{\rm anti}^{\nu}(x_{2}) ] = 0.
\label{3}
\ee
There are a lot of terms with the factor
$
\partial^{\nu}D(x_{1} - x_{2})
$
contributing to the anomaly. There is another subtlety: in the right hand side we have to use in fact the 
chronological products renormalized according to (\ref{r1}) and (\ref{r2}). This brings additional contributions
to the anomaly:
\be
i~{\partial \over \partial x^{\nu}_{1}}~[ \delta(x_{1} - x_{2}) N^{[\mu\nu]\emptyset}(x_{1}) ]
- i~{\partial \over \partial x^{\nu}_{2}}~[ \delta(x_{1} - x_{2}) N^{[\mu][\nu]}(x_{1}) ].
\ee

We must use all the bilinear identities (\ref{Jacobi}) - (\ref{tensor}) from the end of Subsection \ref{ym} and end up with an anomaly 
of the type
\be
C^{\mu}(x_{1}, x_{2}) = \delta(x_{1} - x_{2})~c^{\mu}(x_{1})
\ee
where the expression for 
$
c^{\mu}
$
is rather complicated. We try to equate it with an expression of the type 
$
d^{\rm anti}_{Q}N^{\mu}
$
where
$
N^{\mu}
$
generic form (\ref{r3}). We end up with a system of four equations: from three of them the expressions of
$
g^{(j)}, j = 1,2,3
$
from the statement are derived and the fourth is an identity if we use the bilinear identities (\ref{Jacobi}) - (\ref{tensor}).
So, this anomaly can, again, be eliminated by a finite renormalization (\ref{r3}).

(iv) This repetitive process ends with the consideration of the identity:
\be
d^{\rm anti}_{Q}[ T(x_{1}), T(x_{2})) ] 
- i~{\partial \over \partial x^{\mu}_{1}}~[ T_{\rm anti}^{\mu}(x_{1}), T(x_{2})]
- i~{\partial \over \partial x^{\mu}_{2}}~[ T(x_{1}), T_{\rm anti}^{\mu}(x_{2}) ] = 0.
\label{4}
\ee
As before, we select the terms from the commutators with the factor
$
\partial^{\mu}D(x_{1} - x_{2}).
$
They will produce a piece of the anomaly after the substitution (\ref{com-02}) $\rightarrow$ (\ref{chr-02}). There is another
piece coming from the finite renormalization (\ref{r3}):
\be
i~{\partial \over \partial x^{\mu}_{1}}~[ \delta(x_{1} - x_{2}) N^{[\mu]\emptyset}(x_{1}) ] + (x_{1} \longleftrightarrow x_{2}).
\ee
If we use the bilinear identities  (\ref{Jacobi}) - (\ref{tensor}) from the end of Subsection \ref{ym} and end up with an anomaly 
of the type
\be
D(x_{1}, x_{2}) = \delta(x_{1} - x_{2})~d(x_{1})
\ee
where the expression for 
$
d
$
is complicated. We try to equate it with an expression of the type 
$
d^{\rm anti}_{Q}N
$
where
$
N
$
generic form (\ref{r4}). We end up with a system of four equations: from three of them the expressions of
$
h^{(j)}, j = 1,2,3
$
from the statement are derived and the fourth is an identity if we use the bilinear identities (\ref{Jacobi}) - (\ref{tensor}).
So, this anomaly can, again, be eliminated by a finite renormalization (\ref{r4}).
$\qed$

We end with an explanation of the fact that we obtain nothing new from the anti-BRST transform. In fact, anti-BRST is equivalent to
BRST transform. One can see this if replaces the interaction Lagrangian $T$ from (\ref{T}) by an equivalent Lagrangian i.e. an expression
$
T^{\prime}
$
differing from $T$ by a co-boundary. In fact, we can rewrite the second term from (\ref{T}) in the following way:
\be
T_{2} = {1\over 2}~f_{abc} \left( u_{a}~v_{b}^{\mu}~\partial_{\mu}\tilde{u}_{c} 
- \partial_{\mu}u_{a}~v_{b}^{\mu}~\tilde{u}_{c} - m_{a}~\Phi_{a}~\tilde{u}_{b}~u_{c}\right)
+ d_{Q}B + \partial_{\mu}B^{\mu}
\ee
where we omit the explicit expressions $B$ and 
$
B^{\mu}.
$
It is known that we can discard the co-boundary
$
d_{Q}B + \partial_{\mu}B^{\mu}
$
without modifying the values of scattering matrix taken in the subspace of the physical states \cite{trivial}. We can prove that 
the new interaction Lagrangian obtained in this way is anti-symmetric with respect to the transform (\ref{uu}). This
anti-symmetry property can be extended to higher order chronological products, so if we have gauge invariance with respect to
the BRST transform, then we must have gauge invariance with respect to the anti-BRST transform also.

\section{Conclusions}

The main point of this note is that the elimination of the anomalies in higher orders of the perturbation theory is an 
extremely difficult problem. All simple ideas, like for instance, the use of new symmetries, as the anti-BRST symmetry, do
not produce new constraints on the anomalies.

\end{document}